\documentclass[prl,twocolumn,preprintnumbers,amsmath,amssymb,unsortedaddress,superscriptaddress]{revtex4}
\usepackage{graphicx}
\usepackage{dcolumn,color}
\usepackage{bm}
\usepackage{overpic}
\usepackage{multirow}
\usepackage{epsfig}
\usepackage[english]{babel}

\begin{document}
\bibliographystyle{try}

\newcounter{univ_counter}
\setcounter{univ_counter} {0} \addtocounter{univ_counter} {1}
 \edef\HISKP{$^{\arabic{univ_counter}}$ }\addtocounter{univ_counter}{1}
 \edef\PI{$^{\arabic{univ_counter}}$ } \addtocounter{univ_counter}{1}
 \edef\GATCHINA{$^{\arabic{univ_counter}}$ }\addtocounter{univ_counter}{1}
 \edef\BOCHUM{$^{\arabic{univ_counter}}$ } \addtocounter{univ_counter}{1}
 \edef\GIESSEN{$^{\arabic{univ_counter}}$ } \addtocounter{univ_counter}{1}
 \edef\FSU{$^{\arabic{univ_counter}}$ } \addtocounter{univ_counter}{1}
 \edef\BASEL{$^{\arabic{univ_counter}}$ } \addtocounter{univ_counter}{1}

\title{\boldmath The $N(1520)\,3/2^-$ helicity amplitudes from an energy-independent multipole
analysis based on new polarization data on photoproduction of neutral
    pions
}


\affiliation{Helmholtz--Institut f\"ur Strahlen-- und Kernphysik, Universit\"at Bonn, 53115 Bonn, Germany}
\affiliation{Physikalisches Institut, Universit\"at Bonn, 53115 Bonn, Germany}
\affiliation{Petersburg Nuclear Physics Institute, Gatchina, 188300 Russia}
\affiliation{Institut f\"ur Experimentalphysik I, Ruhr--Universit\"at Bochum, 44780  Bochum, Germany}
\affiliation{II.~Physikalisches Institut, Universit\"at Gie{\ss}en, 35392 Gie{\ss}en, Germany}
\affiliation{Department of Physics, Florida State University, Tallahassee, FL 32306, USA}
\affiliation{Department Physik, Universit\"at Basel, 4056 Basel, Switzerland}

\author{J.~Hartmann\HISKP}
\author{H.~Dutz\PI}
\author{A.V.~Anisovich\mbox{\HISKP\hspace{-1mm}$^,$\GATCHINA}}
\author{D.~Bayadilov\mbox{\HISKP\hspace{-1mm}$^,$\GATCHINA}}
\author{R.~Beck\HISKP}
\author{M.~Becker\HISKP}
\author{Y.~Beloglazov\GATCHINA}
\author{A.~Berlin\BOCHUM}
\author{M.~Bichow\BOCHUM}
\author{S.~B\"ose\HISKP}
\author{K.-Th.~Brinkmann\HISKP\hspace{-1mm}$^,$\GIESSEN}
\author{V.~Crede\FSU}
\author{M.~Dieterle\BASEL}
\author{H.~Eberhardt\PI}
\author{D.~Elsner\PI}
\author{K.~Fornet-Ponse\PI}
\author{St.~Friedrich\GIESSEN}
\author{F.~Frommberger\PI}
\author{Ch.~Funke\HISKP}
\author{M.~Gottschall\HISKP}
\author{A.~Gridnev\GATCHINA}
\author{M.~Gr\"uner\HISKP}
\author{E.~Gutz\mbox{\HISKP\hspace{-1mm}$^,$\GIESSEN}}
\author{Ch.~Hammann\HISKP}
\author{J.~Hannappel\PI}
\author{V.~Hannen\GIESSEN}
\author{J.~Herick\BOCHUM}
\author{W.~Hillert\PI}
\author{Ph.~Hoffmeister\HISKP}
\author{Ch.~Honisch\HISKP}
\author{O.~Jahn\PI}
\author{T.~Jude\PI}
\author{A.~K\"aser\BASEL}
\author{D.~Kaiser\HISKP}
\author{H.~Kalinowsky\HISKP}
\author{F.~Kalischewski\HISKP}
\author{P.~Klassen\HISKP}
\author{I.~Keshelashvili\BASEL}
\author{F.~Klein\PI}
\author{E.~Klempt\HISKP}
\author{K.~Koop\HISKP}
\author{B.~Krusche\BASEL}
\author{M.~Kube\HISKP}\author{M.~Lang\HISKP}
\author{I.~Lopatin\GATCHINA}
\author{K.~Makonyi\GIESSEN}
\author{F.~Messi\PI}
\author{V.~Metag\GIESSEN}
\author{W.~Meyer\BOCHUM}
\author{J.~M\"uller\HISKP}
\author{M.~Nanova\GIESSEN}
\author{V.~Nikonov\HISKP\hspace{-1mm}$^,$\GATCHINA}
\author{D.~Novinski\GATCHINA}
\author{R.~Novotny\GIESSEN}
\author{D.~Piontek\HISKP}
\author{C.~Rosenbaum\HISKP}
\author{B.~Roth\BOCHUM}
\author{G.~Reicherz\BOCHUM}
\author{T.~Rostomyan\BASEL}
\author{A.~Sarantsev\mbox{\HISKP\hspace{-1mm}$^,$\GATCHINA}}
\author{Ch.~Schmidt\HISKP}
\author{H.~Schmieden\PI}
\author{R.~Schmitz\HISKP}
\author{T.~Seifen\HISKP}
\author{V.~Sokhoyan\HISKP}
\author{Ph.~Th\"amer\HISKP}
\author{A.~Thiel\HISKP}
\author{U.~Thoma\HISKP}
\author{M.~Urban\HISKP}
\author{H.~van~Pee\HISKP}
\author{D.~Walther\HISKP}
\author{Ch.~Wendel\HISKP}
\author{U.~Wiedner\BOCHUM}
\author{A.~Wilson\HISKP\hspace{-1mm}$^,$\FSU}
\author{A.~Winnebeck\HISKP}
\author{L.~Witthauer\BASEL}
\author{Y.~Wunderlich\HISKP\hspace{-1.5mm}.}

\collaboration{The CBELSA/TAPS Collaboration}\noaffiliation

\date{\today}
\begin{abstract}
New data on the polarization observables  $T$, $P$, and $H$ for the
reaction $\gamma p\to p\pi^0$ are reported. The results are
extracted from azimuthal asymmetries when a transversely polarized
butanol target and a linearly polarized photon beam are used. The data
were taken at the Bonn electron stretcher accelerator ELSA using the
CBELSA/TAPS detector. These and earlier data are used to perform a
truncated energy-independent partial wave analysis in sliced-energy
bins. This energy-independent analysis is compared to the results
from energy-dependent partial wave analyses.
\vspace{2mm}
\end{abstract}


\maketitle

It is more than 50 years ago that Chew, Goldberger, Low, and Nambu
(CGLN) \cite{Chew:1957tf} wrote down the four (complex) amplitudes
governing a seemingly simple process in which single pseudoscalar
mesons, e.g. pions, are produced off protons or neutrons by photons
in the GeV energy range. These four CGLN
amplitudes can be expanded into Legendre polynomials and the
photoproduction multipoles emerge. The multipoles contain the
information on resonances and their properties in a given partial
wave. This information can then be extracted in an energy-dependent
fit to the multipoles.  {\it At least} eight carefully
chosen experiments are required to determine the CGLN amplitudes (up to one
arbitrary phase for each bin in energy and angle) in a
{\it complete} experiment \cite{Barker:1975bp,Chiang:1996em}. In 
practice, a significantly larger number of observables need 
to be known when limitations in statistics and accuracy 
of experimental data are taken into account \cite{Sandorfi:2010uv}.  
A direct fit to the data with a truncated series of multipoles 
is certainly more realistic. In the region below the 2$\pi$ threshold, $S$
and $P$ waves are sufficient to describe $\pi^0$ photoproduction,
and a measurement of differential
cross sections $d\sigma/d\Omega$ and the photon beam asymmetry
$\Sigma$ is sufficient to determine the contributing multipoles
\cite{Beck:1997ew,Blanpied:1997zz}. A minimum of five observables is
claimed to be required if the analysis is extended
to include higher waves \cite{Omelaenko1981,Wunderlich:2013iga}.
However, ambiguities may (and will) increase the number
of needed observables. The hope is that photoproduction will
overcome the limitations of pion-induced reactions
\cite{Isgur:2000ad} and provide the information to uncover nucleon
and $\Delta$ resonances predicted by quark models (see e.g.
\cite{Capstick:1986bm,Loring:2001kx,Santopinto:2012nq}) - and now in QCD calculations
on a lattice \cite{Edwards:2011jj} - but not found in experiments
performed with pion beams. This program requires high-intensity beams of photons up to a few
GeV energy with linear and circular polarization, polarized proton
and/or neutron targets, and detection of the polarization of the
outgoing nucleon. These technical requirements are now all met for
a few years, and precise new data including the measurement of double
polarization observables start to be published. But still, an unambiguous
determination of the four CGLN amplitudes is not yet possible.

In this letter, we present new data on three polarization observables
for the reaction
\begin{equation}
\gamma p\to p\pi^0,
\label{reac}
\end{equation}
thus providing an important next step towards the complete
experiment. The observables are the target asymmetry $T$, the proton
recoil asymmetry $P$, and $H$, a double polarization observable
describing the correlation between beam and target asymmetries.
Together with the differential cross section
(e.g.\cite{vanPee:2007tw}) and the data on $\Sigma$
\cite{Bartalini:2005wx,Elsner:2008sn,Sparks:2010vb}, $G$
\cite{Thiel:2012yj} and on $E$ \cite{Gottschall:2013} for
this reaction, seven observables have been determined. One might
therefore expect that a model-independent construction of
photoproduction multipoles for $S$, $P$, and $D$ waves should be
possible. The contributions of higher multipoles are expected to be
small below $W=1600$~MeV; they are approximated by the
energy-dependent Bonn-Gatchina\,(BnGa) fit to a large data base of pion and
photo-induced reactions \cite{Anisovich:2011fc}.

The energy range for which these seven observables exist covers the
$N(1520)$ resonance with spin and parity $J^P=3/2^-$ which decays
into $p\pi^0$ with $L=2$. From the reconstructed $E_{2^-}$ and
$M_{2^-}$ multipoles we deduce the $N(1520)\,3/2^-$ photocouplings. It is
the first time that data are available which allow for an
energy-independent reconstruction of multipoles in the energy range
covering the second resonance region ($N(1520)\,3/2^-$ and 
$N(1535)\,1/2^-$).

The experiment was performed at the Bonn ELectron Stretcher
Accelerator ELSA \cite{Hillert:2006yb}. Bremsstrahlung photons from
a 3.2~GeV electron beam were scattered off a diamond crystal to produce
a linearly polarized photon beam \cite{Elsner:2008sn}. Two
orthogonal settings of the polarization plane were used (called
$\parallel$ and $\perp$). The polarization reached its maximum of
${\rm p_\gamma}=65$\% at 850~MeV and dropped down to 40\% at 700~MeV.
The polarized photon beam impinged on a butanol (${\rm C_4H_{9}OH}$)
target with transversely polarized protons~\cite{Dutz:2004zz}. The
mean proton polarization was ${\rm p_T}\approx75$\%. Data were taken
with two opposite settings of the target polarization direction,
defined as $\uparrow$ and $\downarrow$.

Neutral pions from the reaction~(\ref{reac}) (or from C/O nuclei) were 
reconstructed from their $\gamma\gamma$ decay using the CBELSA/TAPS
electromagnetic calorimeters. They consist of 1320 CsI(Tl)
\cite{Aker:1992ny} and 216 forward BaF$_2$ \cite{TAPS} crystals with
a polar angle coverage down to 1$^\circ$ in forward direction. 
Protons from~(\ref{reac}) were detected in the calorimeters as well.
In the analysis, events with three distinct calorimeter hits were
selected. First they were treated as photon candidates, three
$\gamma\gamma$ invariant masses were formed and a cut on the
$\gamma\gamma$ invariant mass was applied. Then, with the remaining
calorimeter hit as the proton candidate, additional cuts to ensure
momentum conservation were applied. The resulting event sample contains
$1.4$ million $p\pi^0$ events with a background contribution of less
than 1\% in all energy and angular bins. 

The butanol target contained unpolarized C and O nuclei. The dilution
factor $d$ takes into account that photoproduction of $\pi^0$ off
nucleons in C or O nuclei cannot be discriminated completely against
reaction~(\ref{reac}). $d$ is a function of $E_\gamma$ and
$\cos\theta$, it was determined using data for which the butanol
target has been replaced by a carbon foam target inside the cryostat.

In the coordinate frame of the detector we define $\alpha$ as azimuthal
angle of the beam photon polarization plane in the $\parallel$ setting,
$\beta$ as azimuthal angle of the target polarization vector in the
$\uparrow$ setting, and $\phi$ as azimuthal angle of the $\pi^0$.
Then, the differential cross section is modulated according to

\begin{figure}[pt]
\begin{tabular}{cc}
  \hspace{-1mm}\includegraphics[width=0.24\textwidth]{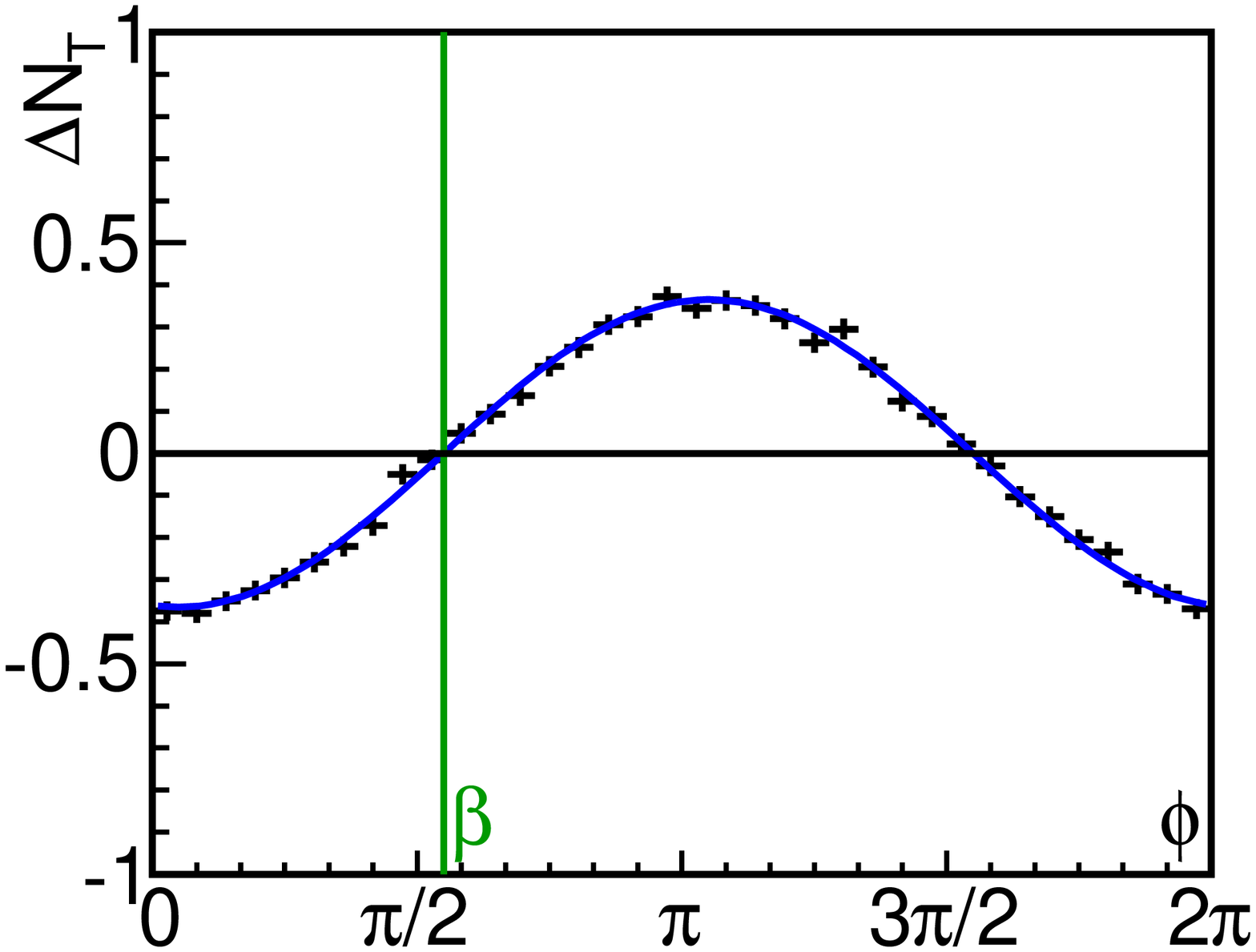}&
\hspace{-1mm}
\includegraphics[width=0.24\textwidth]{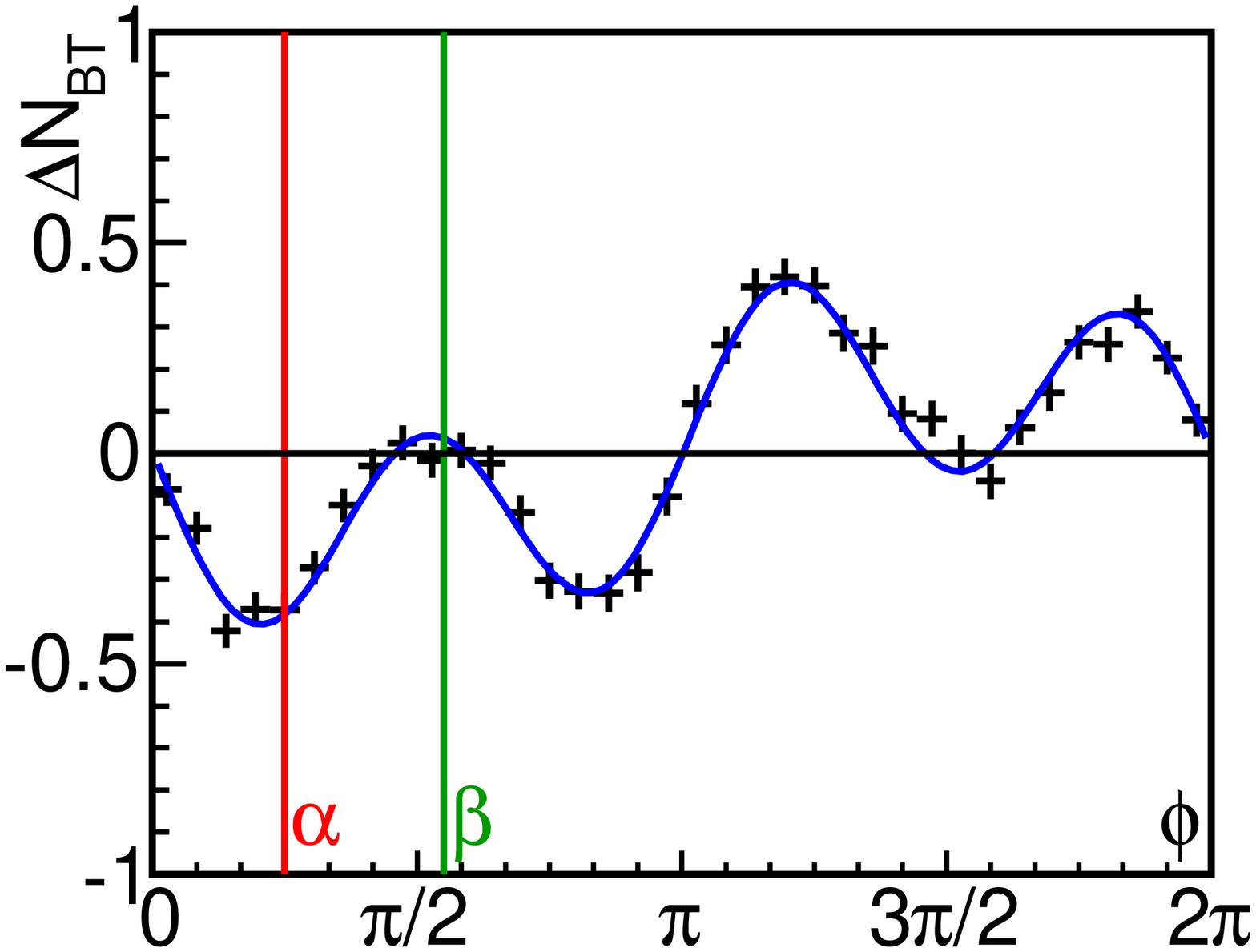}
\end{tabular}
\vspace{-3mm}
\caption{\label{pic:dilution}(Color online) Event yield asymmetry as
a function of $\phi$, left: $\Delta N(\phi)_\text{T}$, right:
$\Delta N(\phi)_\text{\,BT}$, fitted by the functions given in 
eqs.~(\ref{pol-t}) and (\ref{pol-ph}), respectively. Both show data from
the energy bin $W=1.524$\,--\,$1.542$~GeV.}\vspace{-3mm}
\end{figure}

{\footnotesize
\begin{equation}
\hspace{-10mm}\frac{d\sigma}{d\Omega} = \left(
\frac{d\sigma}{d\Omega}\right)_{\!\!0} \cdot \bigl\{ 1-{\rm
p_{\gamma}}\Sigma\cos(2(\alpha-\phi)) + {\rm p_T} T
\sin(\beta-\phi)\vspace{-1mm}
\end{equation}\begin{equation}- {\rm p_{\gamma}} {\rm p_T}
P\cos(2(\alpha-\phi))\sin(\beta-\phi) + {\rm p_{\gamma}} {\rm p_T}
H\sin(2(\alpha-\phi))\cos(\beta-\phi) \bigr\}. \nonumber
\end{equation}}

Since the detector acceptance is identical for all polarization
settings, the cross section can be replaced by the normalized yield
$N$, and the target asymmetry $T$ is determined from a fit to the
azimuthal yield asymmetry:

 {\footnotesize\begin{equation} \label{pol-t}\Delta N(\phi)_\text{\,T}  =
\frac{1}{d \cdot {\rm p_T}} \cdot
\frac{N_\uparrow-N_\downarrow}{N_\uparrow+N_\downarrow} = T \cdot
\sin(\beta-\phi) ,\end{equation}
\begin{equation} \qquad d(E_\gamma,\theta) = \frac{N_\text{butanol} -
 N_\text{carbon}}{N_\text{butanol}}
\end{equation}}

\begin{figure*}[pt]
\vspace{-2mm}
\hspace{-2mm}\includegraphics[width=0.999\textwidth]{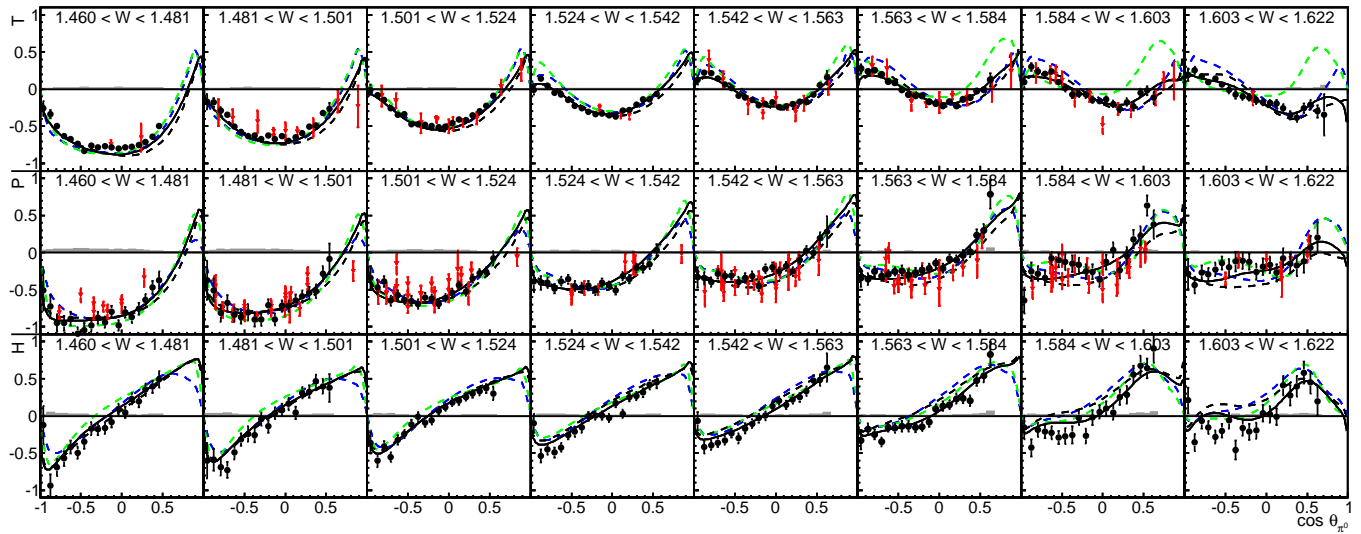}
\vspace{-2mm}\caption{\label{pic:tph}(Color online) The polarization observables
$T$, $P$, and $H$ (energy bins in GeV). References to earlier data (gray triangles, (red)) are given in \cite{Anisovich:2011fc}, refs. 49-71.
The data are compared to predicitions
(dashed curves) from BnGa2011 (black), MAID (light gray, (green)),
and SAID CM12 (dark gray, (blue)). The BnGa refit (BnGa2014) is shown as a black
solid line. Similar fits have been performed within the Bonn-J\"ulich dynamical coupled-channel model \cite{Ronchen:2014cna}.
The systematic errors due to the uncertainty in the degrees of proton (2\%)
and photon (4\%) polarizations, in the dilution factor (1\%-4\%) and the background 
contamination (0.01 absolute error) are shown as a dark gray band. 
An additional systematic error on the photon energy reaching from
$\sigma_{E\gamma}^\text{sys}$ = 6.5~MeV at the lowest to 5.4~MeV
at the highest energy bin plotted is not shown.\vspace{-3mm}}
\end{figure*}

A typical example for such a fit is shown in Fig.~\ref{pic:dilution}, 
left panel. $P$ and $H$ are extracted from data
where not only the target polarization is changed but also the
photon polarization plane from $\parallel$ to $\perp$, using the
equation:

 {\footnotesize\begin{eqnarray}
\label{pol-ph}
\Delta N(\phi)_\text{\,BT} = \frac{1}{d \cdot {\rm p_{\gamma}} {\rm
p_T}} \cdot \frac{(N_{\perp\uparrow}
-N_{\perp\downarrow})-(N_{\parallel\uparrow}-N_{\parallel\downarrow})}
{(N_{\perp\uparrow}+N_{\perp\downarrow})+(N_{\parallel\uparrow}+N_{\parallel\downarrow})} \nonumber\\
= P \sin(\beta-\phi)\cos(2(\alpha-\phi)) - H
\cos(\beta-\phi)\sin(2(\alpha-\phi))\ \
\end{eqnarray}}\noindent
The observables $P$ and $H$ are determined by a
fit to the $\Delta N(\phi)_\text{\,BT}$ distributions, as shown by the example
in Fig.~\ref{pic:dilution}, right.

Fig.~\ref{pic:tph} shows the results for $T$, $P$, and $H$ as
functions of the $\gamma p$ invariant mass $W$. $T$ does not require
a polarized photon beam; hence, results are available up to $W=2.5$~GeV.
The data above $W=1.65$~GeV will be shown elsewhere
\cite{Hartmann:tbp}. All three observables are determined simultaneously.
The results agree well with previously reported
measure\-ments but are more precise and extend the range in both
energy and angles. For the double polarization observable $H$, no
data exist so far below W=1800~MeV. The agreement with predictions
from BnGa2011 \cite{Anisovich:2011fc}, MAID \cite{Drechsel:1998hk},
and SAID (CM12) \cite{Workman:2012jf} is, in general, satisfactory.
Larger differences between the different predictions become visible
e.g. for $T$ at forward angles and higher energies.~A BnGa refit
(solid curve) reproduces the data rather well. The refit includes differential cross sections $d\sigma/d\Omega$ ($\chi^2/N_{\rm data}=8961/5469$)~\cite{GWU}, the beam asymmetry $\Sigma$ (4630/2032) \cite{Bartalini:2005wx,Elsner:2008sn,Sparks:2010vb}, and the double polarization variables $G$ \cite{Thiel:2012yj} and $E$ (1197/827) \cite{Gottschall:2013}.

The full photoproduction amplitude contains contributions from a
series of electric and magnetic multipoles which can be
characterized by the orbital angular momentum $L$ in the decay and
the total spin $J=L\pm1/2$ of the excited wave. We may expect (and
the expectation is supported by partial wave analyses like MAID,
SAID and BnGa) that at low energies higher multipoles contribute
little to the reaction~(\ref{reac}). For an energy-truncated PWA
they could hence be neglected. An improved approach is to use
instead the according multipoles from a model, e.g. from BnGa2014.
Here, we fix all multipoles with $L\ge3$ to those from the BnGa
energy-dependent partial wave analysis while magnitudes and phases
for $E_{0^+}$, $E_{1^+}$, $E_{2^+}$, $E_{2^-}$, $M_{1^+}$,
$M_{2^+}$, and $M_{1^-}$ are left free. One overall phase remains
undetermined, hence we determine the phases relative to the
$M_{2^-}$ phase.\vspace{-0.5mm}

\begin{figure*}
\begin{tabular}{cccccc}
\includegraphics[width=0.167\textwidth]{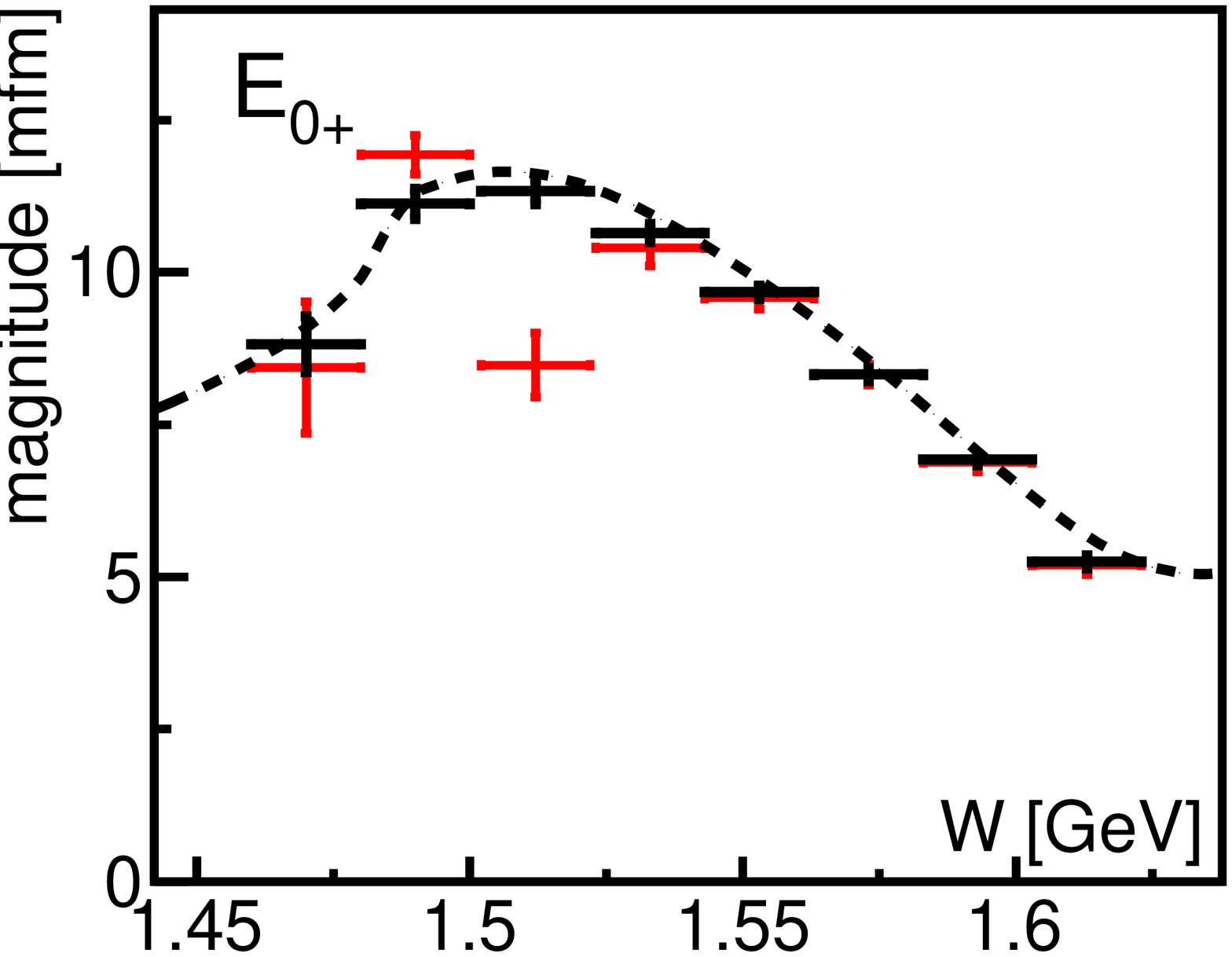}&
\hspace{-2mm}\includegraphics[width=0.167\textwidth]{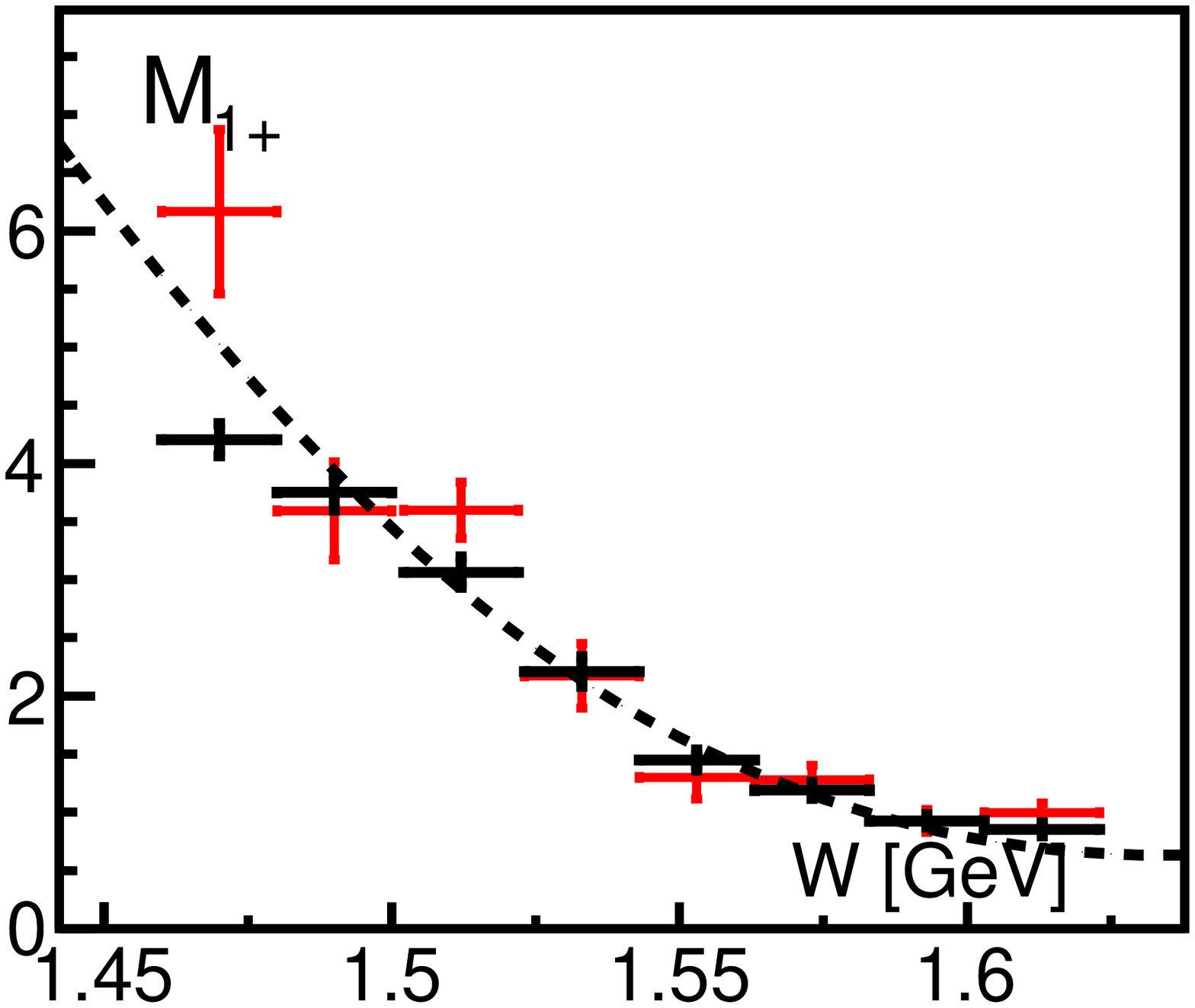}&
\hspace{-2mm}\includegraphics[width=0.167\textwidth]{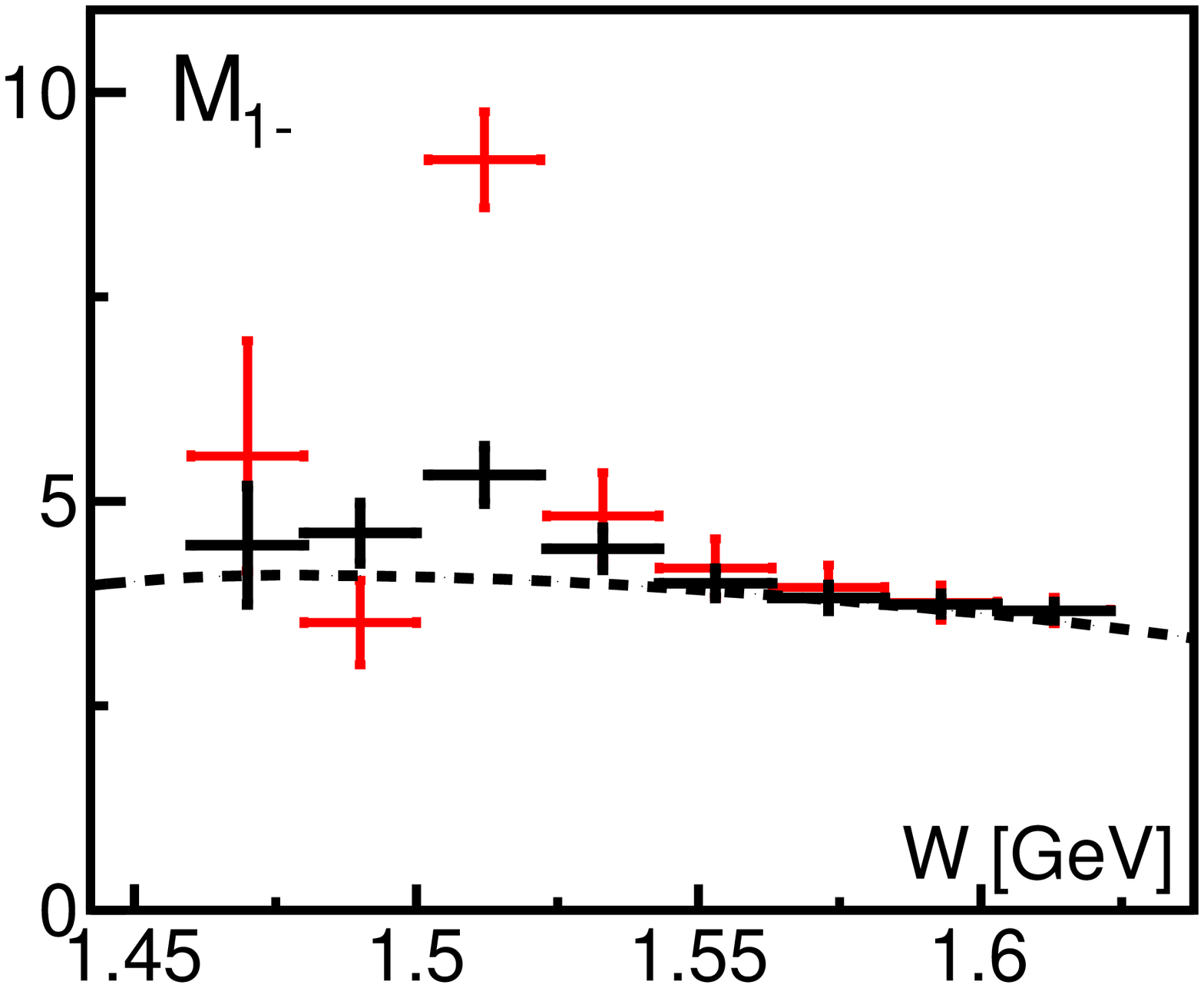}&
\hspace{-2mm}\includegraphics[width=0.167\textwidth]{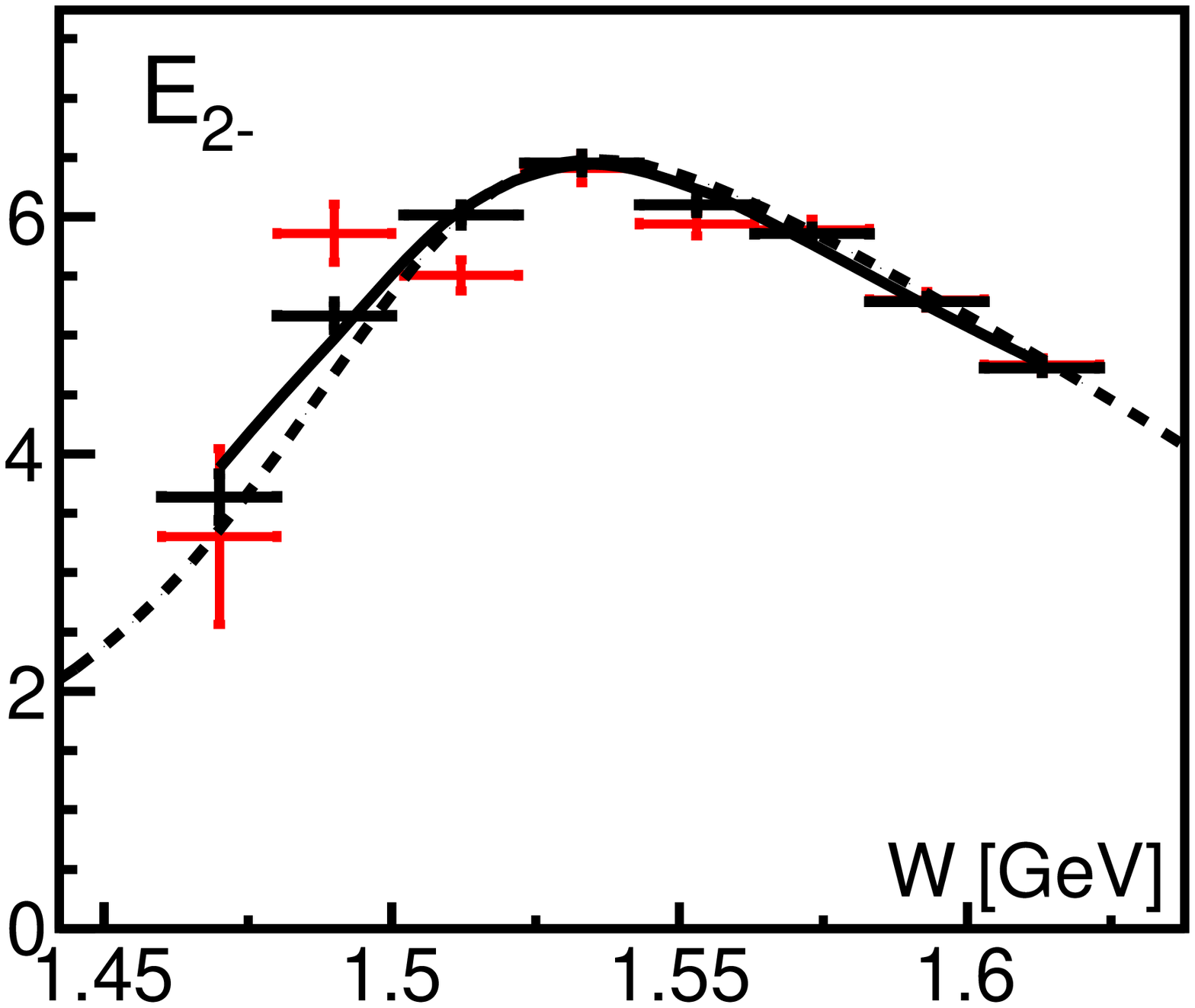}&
\hspace{-2mm}\includegraphics[width=0.167\textwidth]{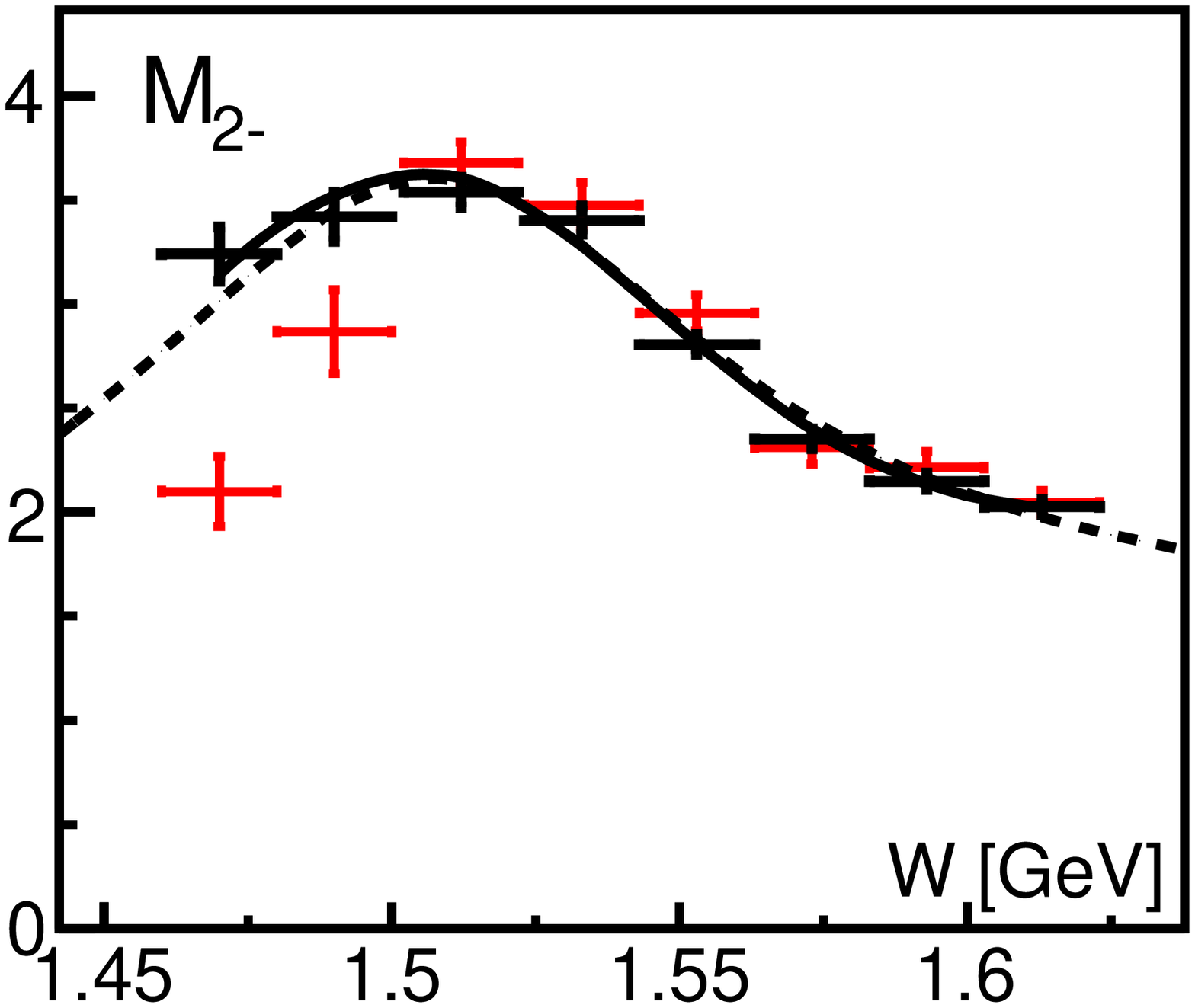}&
\hspace{-2mm}\includegraphics[width=0.167\textwidth]{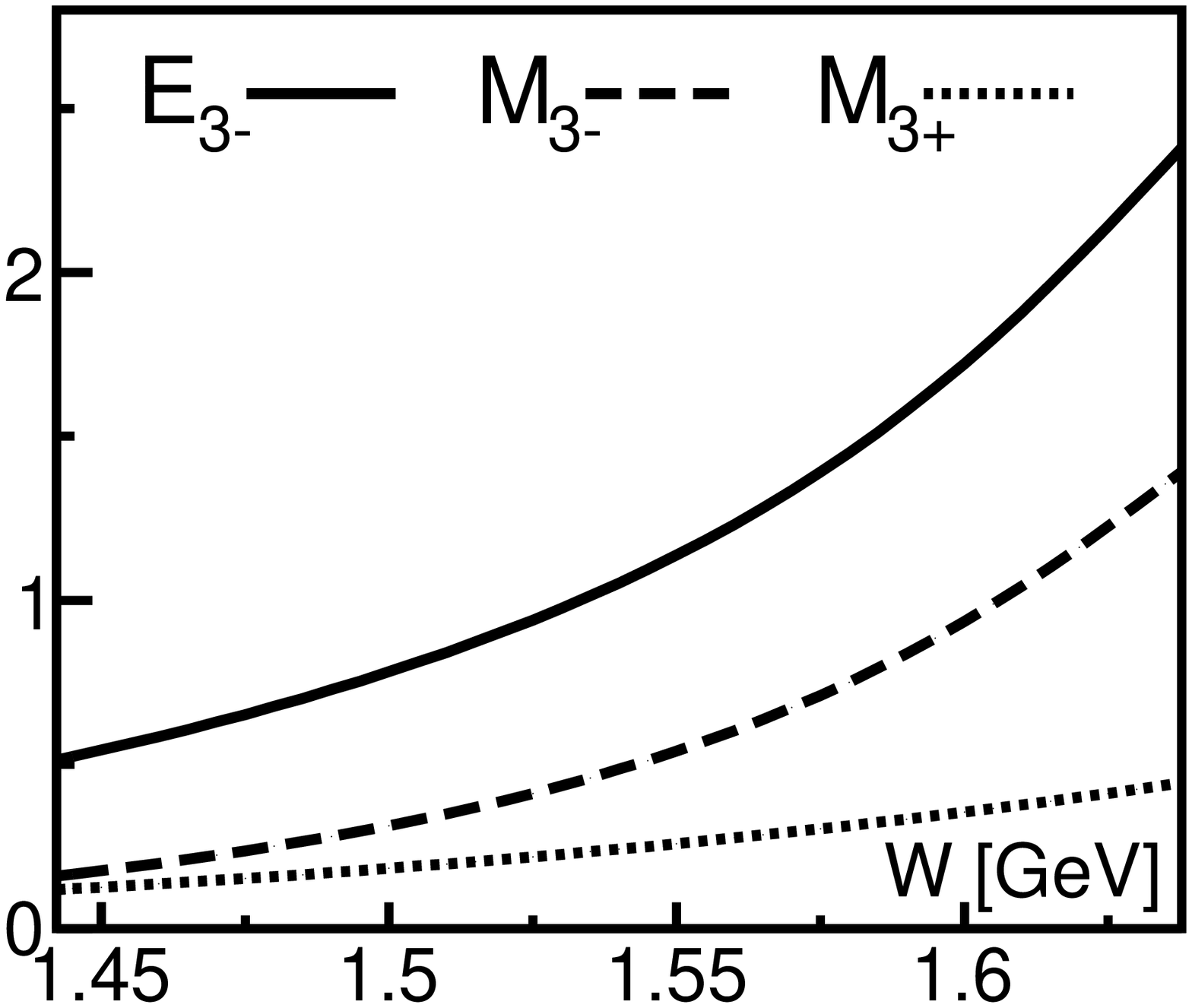}\\
\includegraphics[width=0.167\textwidth]{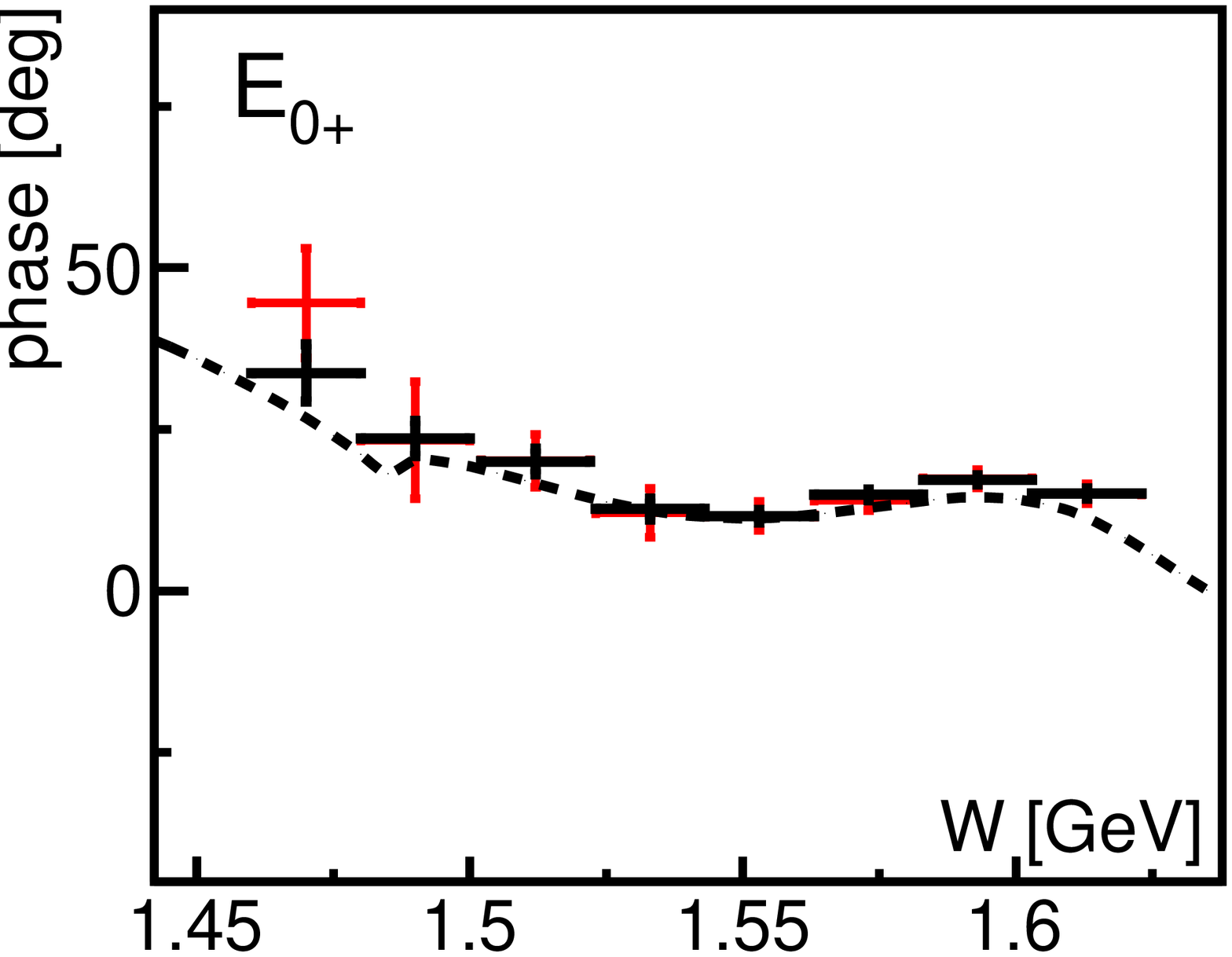}&
\hspace{-2mm}\includegraphics[width=0.167\textwidth]{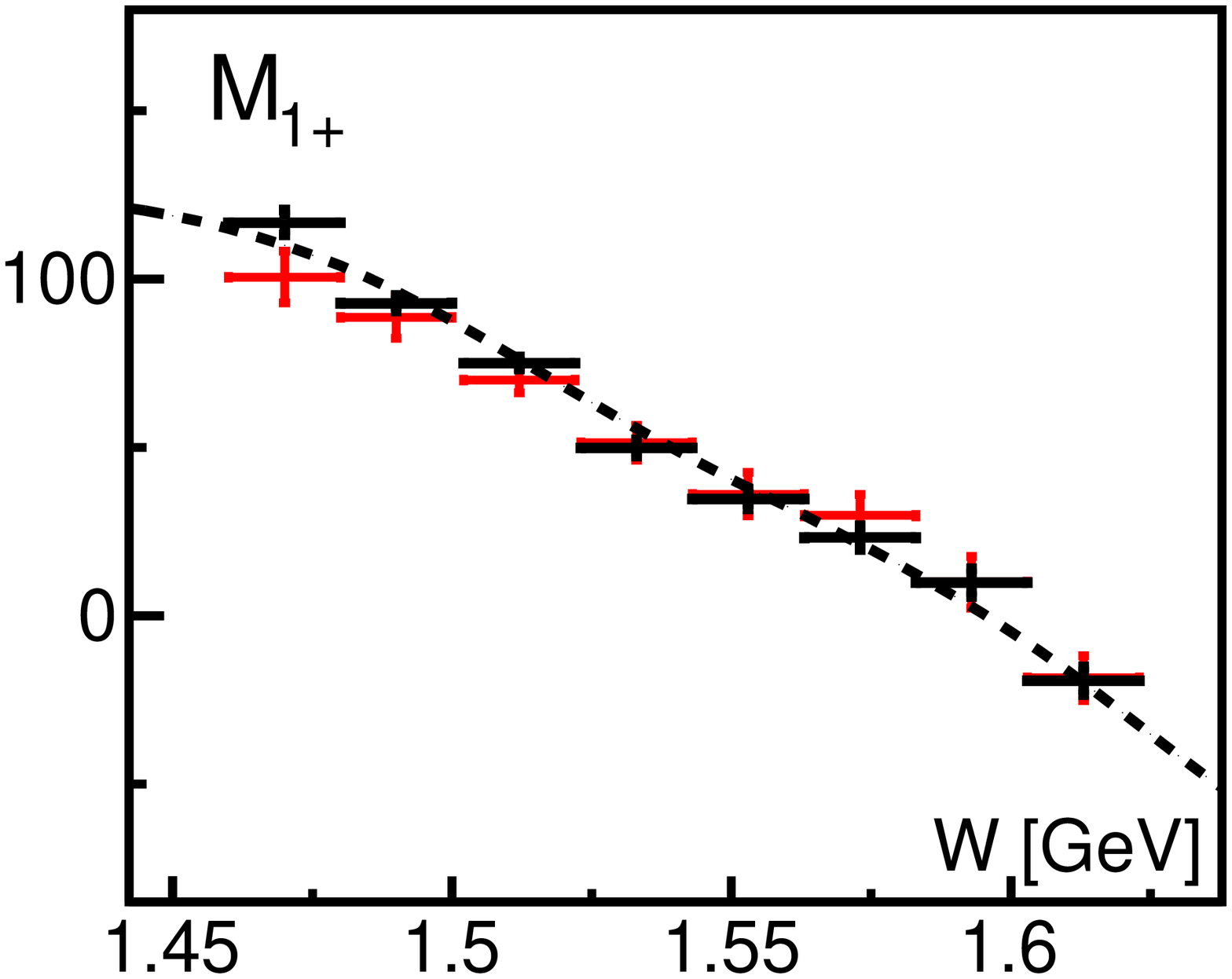}&
\hspace{-2mm}\includegraphics[width=0.167\textwidth]{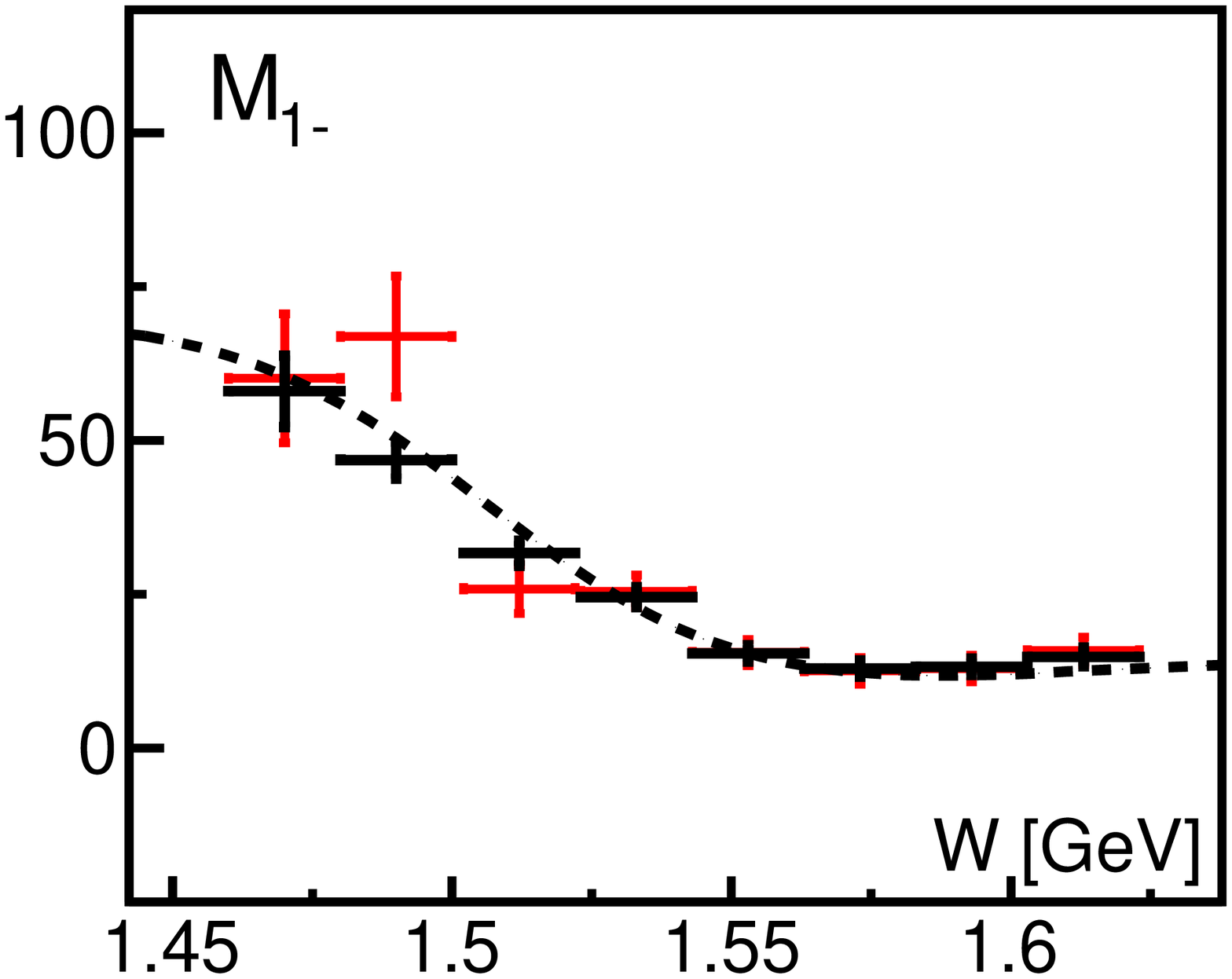}&
\hspace{-2mm}\includegraphics[width=0.167\textwidth]{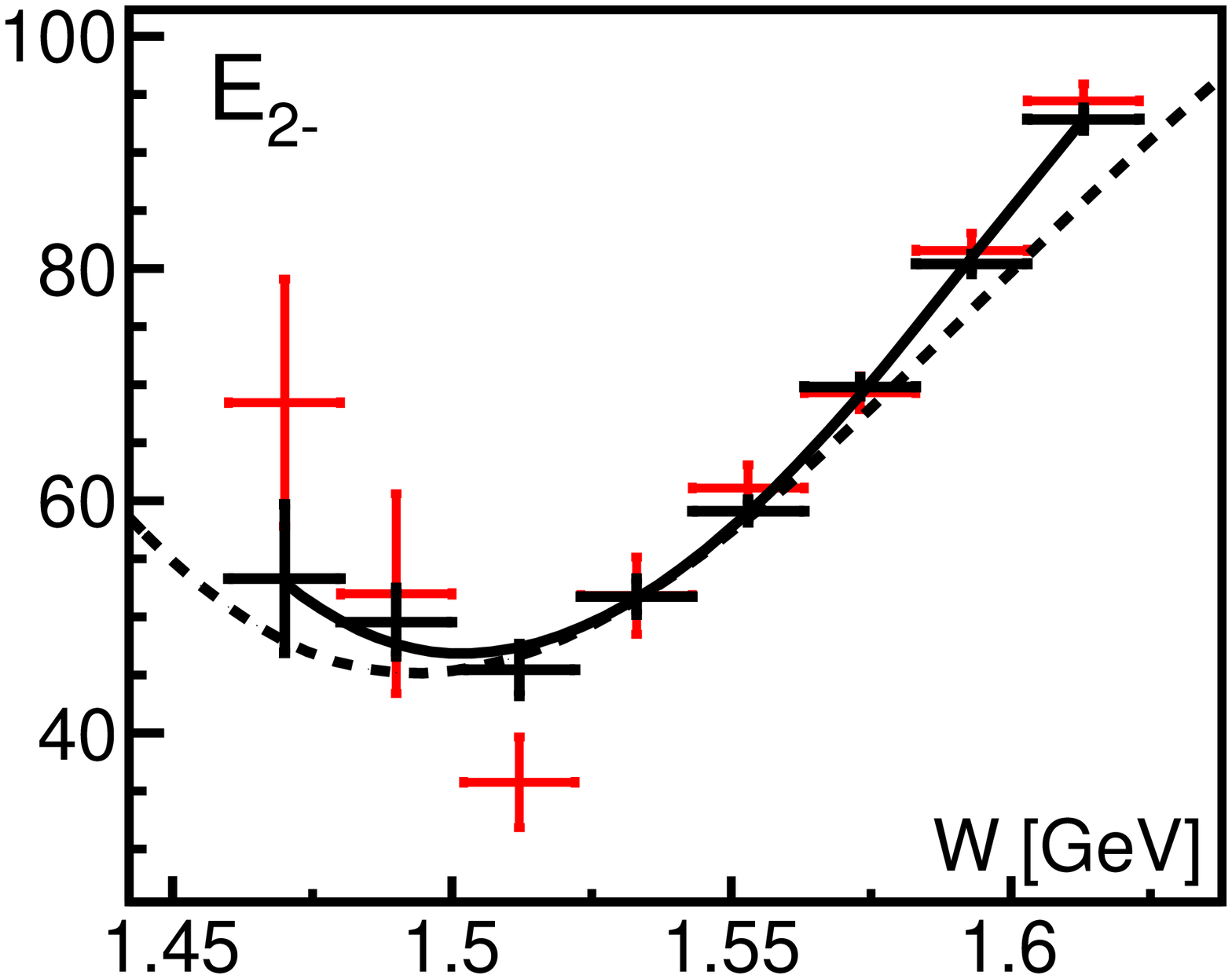}&
\hspace{-2mm}\includegraphics[width=0.167\textwidth]{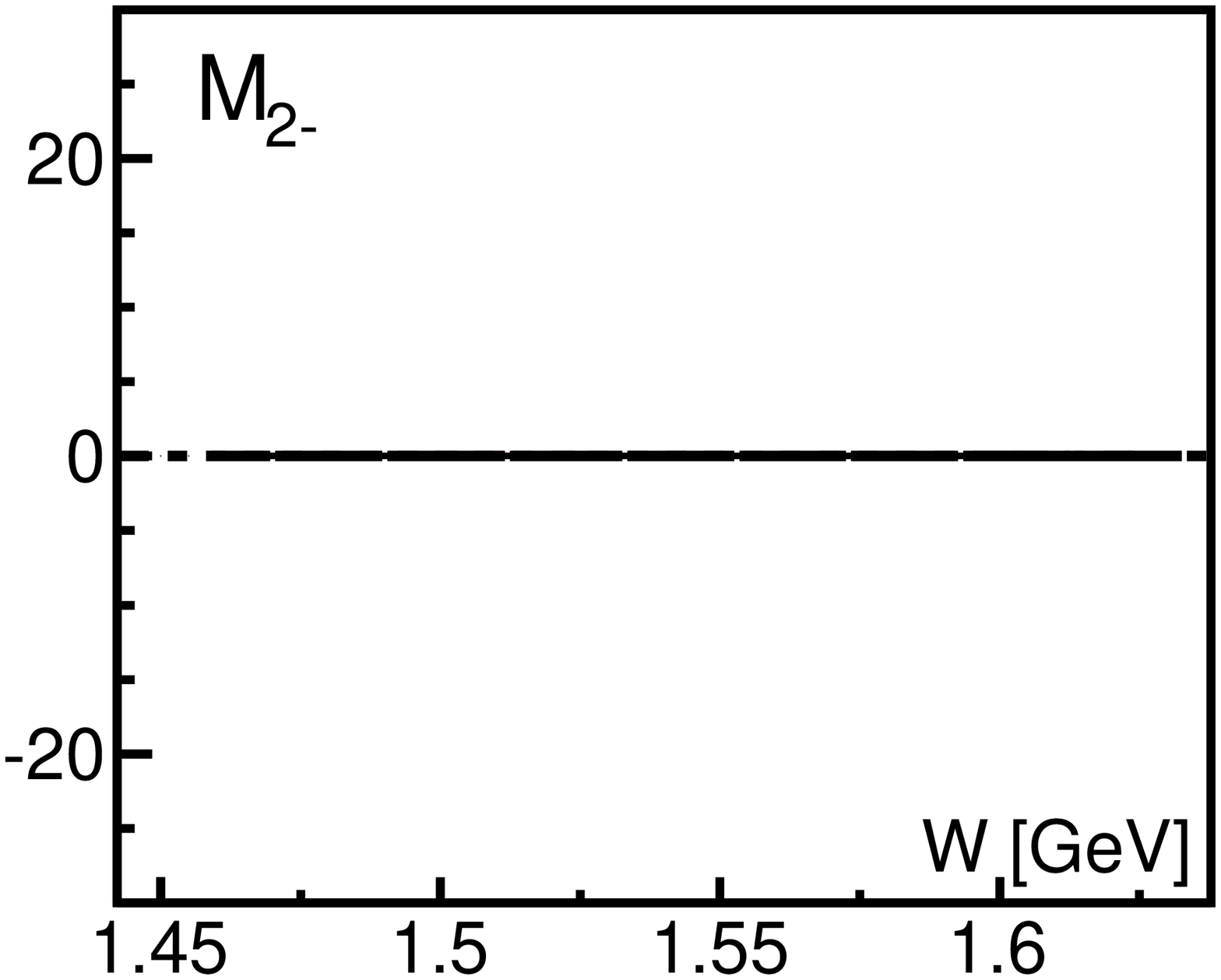}&
\hspace{-2mm}\includegraphics[width=0.167\textwidth]{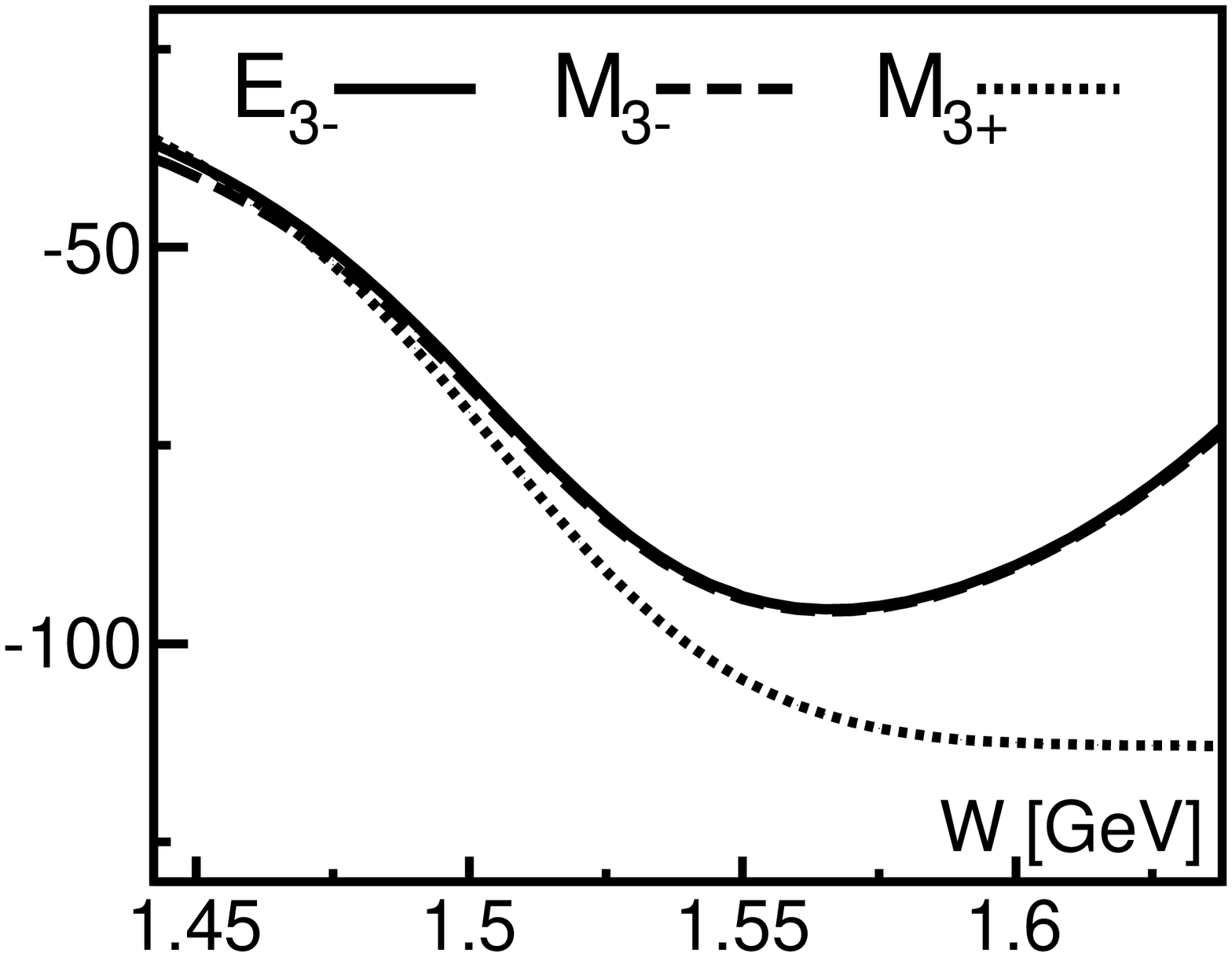}
\end{tabular}
 \caption{\label{edm}(Color online) Magnitude (upper row) and phase (lower row) of
multipoles derived from a fit to data in slices of energy. Grey
(red) crosses show the results of an unbiased fit, black crosses
represent a fit with a penalty function (see text). The dashed lines
show the new energy-dependent Bonn-Gatchina fit (BnGa2014), the
solid lines represent a Breit-Wigner plus background fit to the
black crosses for $E_{2^-}$ and $M_{2^-}$. The largest contributions from higher waves
(from BnGa2014) are shown as well. The $E_{3^-}$ and $M_{3^-}$ multipoles excite 
the close-by resonance $N(1680)5/2^+$, $M_{3^+}$ excites $\Delta(1950)7/2^+$.
\vspace{-2mm}}
\end{figure*}

The  resulting reconstructed multipoles are shown in Fig.~\ref{edm},
except those with magnitudes staying below 1~mfm in the energy
region covered here. The small multipoles scatter around small
values. Note that a factor 10 in the multipole magnitude corresponds
to an intensity ratio of a factor~100.

Most reconstructed multipoles (faint (red) crosses in Fig.~\ref{edm}) are
compatible with the energy-dependent analysis, at least at a
2$\sigma$ level even though a few larger discrepancies
are observed. But deviations in the magnitude are not accompanied by
visible effects in the phase motion; we conclude that these are
artifacts of the fit. Indeed, the fit to the data shows not only one
isolated minimum; instead other local minima exist which describe
angular distributions and polarization observables with similar
quality. To choose among these solutions, we applied a penalty
function adding to the $\chi^2$ of the fit the squared difference
between the reconstructed multipoles (faint (red) crosses in
Fig.~\ref{edm}) and the energy-dependent curve divided by the
corresponding statistical error squared. This penalty has hardly any visible
impact on the fit to $P,T$ and $H$. The resulting reconstructed
multipoles -- shown as black points with error bars in
Fig.~\ref{edm} -- are now fully compatible with the energy-dependent
fit.

\begin{table}[pt]
\vspace{-2mm}
\caption{\label{heli}The $N(1520)\,3/2^-$ helicity amplitudes (in GeV$^{-1/2}$).}
\begin{scriptsize}
\renewcommand{\arraystretch}{1.3}
\begin{tabular}{crrrrr}\hline\hline
$N(1520)\,3/2^-$ 	& this work  &CM12 \cite{Workman:2012jf}& SN11 \cite{Workman:2011vb}& BnGa \cite{Anisovich:2011fc}&PDG \cite{Nakamura:2010zzi}\\\hline
	$A_{1/2}$ &\ $-$0.022  &\ $-$0.019  &\ $-$0.016  &\ $-$0.022 &\ $-$0.024\\[-1ex]
	          &$\!\pm\!$0.009&$\!\pm\!$0.002&$\!\pm\!$0.002&$\!\pm\!$0.004&$\!\pm\!$0.009\\
  $A_{3/2}$ &\  0.118      &0.141         &\  0.156      &\  0.131      &\  0.166 \\[-1ex]
            &$\!\pm\!$0.021\ &$\!\pm\!$0.002\ &$\!\pm\!$0.002\ &$\!\pm\!$0.010&\ $\!\pm\!$0.005\\
\hline\hline 
\end{tabular}
\vspace{-3mm}\end{scriptsize}
\end{table}

The reconstructed $E_{2^-}$ and $M_{2^-}$ multipoles receive
contributions from both isospins $I$; a separation into $I=1/2$ and
$I=3/2$ contributions is -- at present -- not possible due to lack
of polarization data from the charge-related reaction $\gamma p\to
n\pi^+$. But physics helps here: The $N(1520)\,3/2^-$ resonance is far
from $\Delta(1700)\,3/2^-$; its phase variation in the 1500\,MeV region is smooth.
We fit the $E_{2^-}$ and $M_{2^-}$ magnitudes (solid crosses in Fig.~\ref{edm}) and their
respective phase difference using Breit-Wigner amplitudes together
with a background amplitude. The fit returns the 
$N(1520)\,3/2^-$ helicity couplings (in GeV$^{-1/2}$), see Table~\ref{heli}. 
The errors comprise the statistical and systematic errors added quadratically.
The statistical error for $A_{1/2}$ is 0.006, and 0.010 for $A_{3/2}$. These errors include 
those contributing to the error band in Fig.~\ref{pic:tph}. The systematic error receives contribution from several sources.\\
i) The photon energy has an uncertainty of about $\pm$6\,MeV. We combined data 
on $T$, $P$, $H$ with data on $d\sigma/d\Omega$, $\Sigma$, $E$, $G$ with relative 
energy shifts of 0, $\pm5$, and $\pm$10\,MeV, and found no evidence for any systematic 
shifts but an additional spread of the results. The spread is taken as additional uncertainty. It amounts to $0.005$ for $A_{1/2}$ and to $0.015$ for $A_{3/2}$ (in GeV$^{-1/2}$). 
ii)~The background amplitude is assumed to be a constant, linear, or quadratic function in $s$ and/or to be given by the $\Delta(1700)3/2^-$ amplitude of the energy-dependent fit. The results using different background parameterizations are consistent, their spread is used to define a systematic error, $0.005$ for $A_{1/2}$ and $0.011$ for $A_{3/2}$. 
iii) We use a $N(1520)\,3/2^- \to N\pi$ branching ratio of $0.63\pm 0.03$. Its uncertainty is a further systematic error. 

In Table~\ref{heli} we compare our values for the $N(1520)\,3/2^-$ helicity amplitudes with values reported elsewhere. Early results are summarized in PDG 2010 \cite{Nakamura:2010zzi}. In particular $A_{3/2}$ was reported at significantly larger values. The results using the model-independent reconstruction of the amplitudes confirms the results of the BnGa energy-dependent analysis \cite{Anisovich:2011fc} but are at variance with the extremely precise results given by
the SAID group \cite{Workman:2012jf,Workman:2011vb}. 
The new result on $A_{1/2}$ is consistent with earlier
determinations. A similar analysis of the $E_{0^+}$ amplitude returns 
a $N(1535)\,1/2^-$ helicity coupling
in the range 0.070 to 0.140\,GeV$^{-1/2}$ depending on the background model.   

In summary, we have reported a measurement of three polarization
observables, $P$, $T$, and $H$, for the reaction $\gamma p\to p\pi^0$.
These new data represent an important step towards a complete
experiment. The data are used to reconstruct multipoles with $L=0$,
$1$ and $2$. No evidence for additional structures
beyond established resonances is found. The helicity amplitudes of
$N(1520)\,3/2^-$ are deduced with minimal model assumptions. The result is 
inconsistent at the level of more than 2$\sigma$ with older (model-dependent) determinations and supports those of the BnGa PWA.   

We thank the technical staff of ELSA and the par\-ti\-ci\-pating
institutions for their invaluable contributions to the success of
the experiment. We acknowledge support from the \textit{Deutsche
Forschungsgemeinschaft} (SFB/TR16) and \textit{Schweizerischer
Nationalfonds}.


\begin{thebibliography}{00}
 \bibitem{Chew:1957tf}
  G.F.~Chew {\it et al.}, 
  Phys.\ Rev.\  {\bf 106}, 1345 (1957).
  \bibitem{Barker:1975bp}
  I.S.~Barker {\it et al.}, 
  Nucl.\ Phys.\ B {\bf 95}, 347 (1975).
\bibitem{Chiang:1996em}
  W.-T.~Chiang, F.~Tabakin,
  Phys.\ Rev.\  {\bf C55}, 2054 (1997).
\bibitem{Sandorfi:2010uv}
A.M.~Sandorfi {\it et al.}, 
  J.\ Phys.\ G {\bf 38}, 053001 (2011).
 \bibitem{Beck:1997ew}
  R.~Beck {\it et al.},
  Phys.\ Rev.\ Lett.\  {\bf 78}, 606 (1997).
\bibitem{Blanpied:1997zz}
  G.~Blanpied {\it et al.},
  Phys.\ Rev.\ Lett.\  {\bf 79}, 4337 (1997).
\bibitem{Omelaenko1981}
  A.S.~Omelaenko,
  Yad. Fiz. {\bf34}, 730 (1981).
\bibitem{Wunderlich:2013iga} 
  Y.~Wunderlich {\it et al.}, 
  arXiv:1312.0245 [nucl-th].
\bibitem{Isgur:2000ad}
  N.~Isgur in Proc. of NSTAR2000, p403, V.D.~Burkert,
L.~Elouadrhiri, J.J.~Kelly, R.C.~Minehart (eds.), 403.
 \bibitem{Capstick:1986bm}
  S.~Capstick and N.~Isgur,
  Phys.\ Rev.\  D {\bf 34}, 2809 (1986).
 \bibitem{Loring:2001kx}
  U.~L\"oring  {\it et al.}, 
  Eur.\ Phys.\ J.\  A {\bf 10}, 395 (2001).
	\bibitem{Santopinto:2012nq} 
  E.~Santopinto and M.M.~Giannini,
  Phys.\ Rev.\ C {\bf 86}, 065202 (2012).
\bibitem{Edwards:2011jj}
  R.G.~Edwards {\it et al.}, 
  Phys.\ Rev.\ D {\bf 84}, 074508 (2011).
  \bibitem{vanPee:2007tw}
  H.~van Pee {\it et al.},
  Eur.\ Phys.\ J.\ A {\bf 31}, 61 (2007).
\bibitem{Bartalini:2005wx}
  O.~Bartalini {\it et al.},
  Eur.\ Phys.\ J.\ A {\bf 26}, 399 (2005).
  \bibitem{Elsner:2008sn}
  D.~Elsner {\it et al.},
  Eur.\ Phys.\ J.\ A {\bf 39}, 373 (2009).
\bibitem{Sparks:2010vb}
  N.~Sparks {\it et al.},
  Phys.\ Rev.\ C {\bf 81}, 065210 (2010).
%
  \bibitem{Thiel:2012yj}
  A.~Thiel {\it et al.},
  Phys.\ Rev.\ Lett.\  {\bf 109}, 102001 (2012).
%
  \bibitem{Gottschall:2013}
  M.~Gottschall {\it et al.},\ Phys.\ Rev.\ Lett.\ {\bf 112}, 012003 (2014)
 \bibitem{Anisovich:2011fc}
  A.V.~Anisovich {\it et al.}, 
  Eur.\ Phys.\ J.\ A {\bf 48}, 15 (2012).
  \bibitem{Hillert:2006yb}
  W.~Hillert,
  Eur.\ Phys.\ J.\  {\bf A28S1}, 139 (2006).
\bibitem{Dutz:2004zz}
  H.~Dutz {\it et al.},
  Phys.\ Rev.\ Lett.\  {\bf 93}, 032003 (2004).
\bibitem{Aker:1992ny}
    E.~Aker {\it et al.},
    Nucl.\ Instr.\ Meth.\ A {\bf 321}, 69 (1992).
\bibitem{TAPS} R. Novotny, IEEE Trans.\ Nucl.\ Sci.\ {\bf NS-38}, 379 (1991).
 \bibitem{Hartmann:tbp}
 J.~Hartmann {\it et al.}, in preparation.
 \bibitem{Drechsel:1998hk}
  D.~Drechsel {\it et al.},
  Nucl.\ Phys.\  {\bf A645}, 145 (1999).
\bibitem{Workman:2012jf}
  R.~L.~Workman {\it et al.}, 
  Phys.\ Rev.\ C {\bf 86}, 015202 (2012).
\bibitem{GWU}W.M.~Briscoe {\it et al.}, 
gwdac.phys.gwu.edu/
\bibitem{Ronchen:2014cna}
  D.~R\"onchen  {\it et al.}, 
 Eur.\ Phys.\ J.\ A {\bf 50}, 101 (2014).
 \bibitem{Workman:2011vb} 
  R.~L.~Workman  {\it et al.}, 
  Phys.\ Rev.\ C {\bf 85}, 025201 (2012).
	\bibitem{Nakamura:2010zzi} 
  K.~Nakamura {\it et al.},
  J.\ Phys.\ G {\bf 37}, 075021 (2010).   

\end{thebibliography}
\end{document}